\documentclass[amssymb,onecolumn]{revtex4-1}
\usepackage{graphicx}
\usepackage[latin1]{inputenc}
\usepackage{epsfig}
\usepackage{amsmath}
\usepackage{amsfonts}
\usepackage{amssymb}

\def\gsim{\:\raisebox{-0.5ex}{$\stackrel{\textstyle>}{\sim}$}\:}

\def\beq{\begin{equation}}
\def\eeq{\end{equation}}
\def\bea{\begin{eqnarray}}
\def\eea{\end{eqnarray}}

\def\nl{\nonumber\\}
\def\roughly#1{\mathrel{\raise.3ex\hbox
{$#1$\kern-.75em\lower1ex\hbox{$\sim$}}}}

\def\gsim{\roughly>}
\def\lesssim{\mathrel{\hbox{\rlap{\hbox{\lower4pt\hbox{$\sim$}}}\hbox{$<$}}}}
\def\gtrsim{\mathrel{\hbox{\rlap{\hbox{\lower4pt\hbox{$\sim$}}}\hbox{$>$}}}}

\def\sla#1{\raise.15ex\hbox{$/$}\kern-.57em #1}

\def\be{\begin{equation}}
\def\ee{\end{equation}}
\def\bea{\begin{eqnarray}}
\def\eea{\end{eqnarray}}
\def\beast{\begin{eqnarray*}}
\def\eeast{\end{eqnarray*}}

 \normalsize

\begin{document}
\begin{flushright}
UMISS-HEP-2012-08
\end{flushright}

\title{Predictions for $B \to \tau \bar{\mu} + \mu \bar{\tau} $}
\author{ Dris Boubaa$^{1,2}$, Alakabha Datta $^{3}$, Murugeswaran Duraisamy$^{3}$, and Shaaban
Khalil$^{1,4,5}$} \affiliation{$^1$Center for Theoretical Physics,
Zewail City for Science and Technology, 6 October City, Cairo, Egypt.\\
$^2$Department of Physics, Faculty of Science, Hassiba Benbouali
University of Chlef, 02000 Chlef, Algeria.
\\
$^3$Department of Physics and Astronomy, 108 Lewis Hall,
University of Mississippi, Oxford, MS 38677-1848, USA.\\
$^4$Department of Mathematics, Faculty of Science,  Ain Shams University, Cairo, Egypt.\\
$^5$School of Physics and Astronomy, University of Southampton,
Highfield, Southampton SO17 1BJ, UK.}

\date{\today}

\begin{abstract}
The observation of $B \to \tau \bar{\mu} + \mu \bar{\tau} $ at present
experiments would be a clear sign of new physics. In this paper we calculate
this process in an extended higgs sector framework where the decay is mediated by the exchange
of spin zero particle with flavour changing neutral current couplings. If we
identify the scalar with the the newly discovered state at LHC with a mass $%
\sim 125 $ GeV then we find that, after imposing all experimental
constraints, the $BR(B_s \to \tau \bar{\mu} + \mu \bar{\tau}) $
can be as high  $\sim 10^{-6}$ and $BR(B_d \to \tau \bar{\mu} +
\mu \bar{\tau})$ can be as high as $\sim 10^{-7}$.  We also
calculate this process in minimal supersymmetric standard model
and find the $BR(B_s \to \tau \bar{\mu} + \mu \bar{\tau})$ is
typically of order $ \sim 10^{-9}$.
\end{abstract}

\maketitle

\affiliation{Department of Physics, Faculty of Science, University
of Chlef, Hay Salem, route nationale 19
 DZ 02000 Chlef, Algeria.}

\affiliation{Dept of Physics and Astronomy, 108 Lewis Hall,
University of Mississippi, Oxford, MS 38677-1848, USA.}

\affiliation{Center for Theoretical Physics, Zewail City for Science and Technology, 6 October City, Cairo, Egypt.\\
School of Physics and Astronomy, University of Southampton,
Highfield, Southampton SO17 1BJ, UK.\\
Department of Mathematics, Ain Shams University, Faculty of
Science, Cairo, 11566, Egypt.}



\section{Introduction}

Flavor changing neutral current (FCNC) events are rare in the Standard Model
(SM) both in the quark and the lepton sectors. These processes can be
affected by new physics (NP). FCNC involving the third family quark and
leptons are interesting as their larger masses make them more susceptible to
new physics effects in various extensions of the SM. As an example, in
certain versions of the two Higgs doublet models (2HDM) the couplings of the
new Higgs bosons are proportional to the masses and so new physics effects
are more pronounced for the heavier generations. Moreover, the constraints
on new physics involving, specially the third generation leptons and quarks,
are somewhat weaker allowing for larger new physics effects. Hence, in the
down quark sector there is intense interest in the search for NP in FCNC  $b
\to s$ and $b \to d$ transitions in $B_{d,s}$ mixing, $b \to s(d) q \bar{q} $
and $b \to s(d) \ell^+ \ell^-$ decays. Many measurements in the $B$ sector
have constrained NP effects though slight discrepancies from SM predictions
still remain in several decays \cite{dattaBnp}.  In the up quark sector one
can look for NP is the processes $t \to c (u)$ transitions \cite{datta_top}.

In the lepton sector FCNC decays are severely suppressed and lepton number
for each family is conserved in the SM. Therefore, FCNC effects and lepton
flavor violation in this sector are sensitive probes of NP. The presence of
neutrino masses and mixing suggest NP in the lepton sector which in turn
could enhance FCNC effects and lepton flavor violation. It is interesting
that measurements involving the the tau lepton show some discrepancies from
SM expectations. The branching ratio of $B \to \tau \nu_{\tau}$ shows some
tension with the SM predictions \cite{belletau} though more recent
measurement are consistent with the SM \cite{new_belletau}. This could
indicate NP \cite{Bhattacherjee:2010ju}, possibly coming from an extended
scalar or gauge sector. There is also a seeming violation of universality in
the tau lepton coupling to the $W$ suggested by the LEP II data which could
indicate new physics associated with the third generation lepton \cite%
{MartinGon}.

More recently, the BaBar collaboration with their full data sample of an
integrated luminosity 426 fb$^{-1}$ has reported the measurements of the
quantities \cite{:2012xj}
\begin{eqnarray}  \label{babarnew}
R(D) &=& \frac{BR(\bar{B} \to D \tau^{-} \bar{\nu_\tau})} {BR(\bar{B} \to D
\ell^{-} \bar{\nu_\ell})}=0.440 \pm 0.058 \pm 0.042\, ,  \notag \\
R(D^*) &=& \frac{BR(\bar{B} \to D^{*} \tau^{-} \bar{\nu_\tau})} {BR(\bar{B}
\to D^{*} \ell^{-} \bar{\nu_\ell})}=0.332 \pm 0.024 \pm 0.018 \, .
\end{eqnarray}
The SM predictions for $R(D)$ and $R(D^*)$ are \cite%
{:2012xj,Fajfer:2012vx,Sakaki:2012ft}
\begin{eqnarray}
R(D) &=& 0.297 \pm 0.017 \, ,  \notag \\
R(D^*) &=& 0.252 \pm 0.003 \,,
\end{eqnarray}
which deviate from the BaBar measurements by 2$\sigma$ and 2.7$\sigma$
respectively. The BaBar collaboration themselves reported a 3.4$\sigma$
deviation from SM when the two measurements of Eq.~\ref{babarnew} are taken
together. This result, if confirmed, could indicate NP involving new charged
scalar or vector boson states \cite{datta_Dtaunu}. Models with new charged
scalars or charged vector bosons also include neutral particles that can
cause FCNC effects.

In this paper we will focus on the process $B_{s,d} \to \ell_i \ell_j$ .
This involves two FCNC interactions- one in the quark and one in the lepton
sector. Typically one believes that this process will be very small even
with new physics and in many NP models this is indeed true. This is because
in these NP models this FCNC process arises in loops and is suppressed. In
this paper we will focus on the decay $B \to \tau \bar{\mu} + \mu \bar{\tau}$
and make predictions for its rate. This decay was considered in earlier
papers in various extensions of the SM \cite{btaumu_old}. In this work we
will calculate the process in two scenarios representing two extension of
the SM.

In the first scenario we will consider this process mediated by a scalar
which we will call $X$ in a model with an extended higgs sector.
In this model we will assume that there are scalar particles that have FCNC couplings and can mediate the decay $ B \to \tau \mu$. We will identify the lowest mass such state with the $X$ particle. It is tue that in this model there can be other heavier neutral states that can also mediate the decay $ B \to \tau \mu$.
When multiple scalar states contribute to the decay there will be more parameters that
cannot be all constrained from the data resulting in loss of predictive power.
To regain some predictive power, our assumption will be that that the dominant contribution to the decay comes from the lowest lying state $X$. We expect that the contributions to the decay from the higher mass states will in general not lower our estimated branching ratios significantly unless there are strong cancellations among the various amplitudes. We will assume that there are no strong cancellations among the various contributions.

We will
identify $X$ with the newly discovered particle, with mass $\sim 125 $ GeV,
observed at LHC \cite{ ATLAS_h, CMS_h} with supporting evidence for its
existence from Fermilab \cite{CDFD0_h}. In this case $B \to \tau \bar{\mu} +
\mu \bar{\tau}$ is generated at tree level due to the presence of $bqX$ and $%
\tau \mu X$ couplings. The coupling $b q X$ can be constrained by $B_{d,s}$
mixing while the $\tau \mu X$ couplings can be constrained from rare $\tau$
decays. Recently in Ref.~\cite{Blankenburg:2012ex,Harnik:2012pb} these
coupling were considered and it was noticed that the $\tau \mu X $ coupling
could be of similar size as the SM higgs to $\tau \tau$ coupling.
 Since $%
B_{s,d}$ are pseudoscalars,  the
flavor changing coupling of $X$ to the $B_{s,d}$ mesons must contain a pseudoscalar
component of the form $(\bar{q} \gamma_5 b)X$ where $q=s,d$.

The spin parity of the $X$ particle is not yet known though various
strategies that allow for the spin and parity determination are being
actively pursued \cite{sanz}. In this scenario the branching ratio for $B
\to \tau \bar{\mu} + \mu \bar{\tau}$ can be significant and can potentially
be observed  at LHCb.
 Generally, final states with $\tau$ leptons are difficult
to measure at LHCb, however because of an associated $\mu$ in the final state
this process may be somewhat easier to observe than $B_{d,s} \to \tau^+ \tau^-$.
 We would like to  point  that the decay can also be mediated through the exchange of a spin one particle at tree level. Since we  will be identifying the mediator with the newly discovered state at LHC we will not consider the mediator to be  a spin one particle. Furthermore, as can be seen from Eq.~\ref{BrBlilj} the matrix elements with scalar currents, associated with a scalar mediator, are enhanced relative to vector currents associated with a spin one mediator.

In the second scenarios we will consider a popular
extension of the SM- the minimal supersymmetric SM (MSSM). Here the FCNC
couplings arise via loops and so our expectation is that the branching ratio
for $B \to \tau \bar{\mu} + \mu \bar{\tau}$ will be tiny. The main point
here is that if this process were to be observed at the LHCb in the near
future this would indicate the presence of a rather non-traditional NP with
appreciable
level FCNC couplings in the lepton sector and possibly a light spin
zero mediator.

The paper is organized in the following manner. In the next section we
describe the general effective Hamiltonian for $b \to sl^-_il^+_j$ decays.
In the following section we consider the decay $B \to \tau \bar{\mu} + \mu
\bar{\tau} $ in  an extended higgs sector model with FCNC higgs couplings. This is followed
by a calculation of $B \to \tau \bar{\mu} + \mu \bar{\tau}$ in MSSM. Finally
we summarize our results and present our conclusions.

\section{Effective Hamiltonian \label{sec2}}

In this section we discuss the effective Hamiltonian for $b \to q(=d,s)
l^-_il^+_j$ transitions. We will focus on the $b \to sl^-_il^+_j$
Hamiltonian as the $b \to d l_i^-l^+_j$ Hamiltonian can be obtained from it
with the obvious replacements.

The effective Hamiltonian for the quark-level transition $b \to sl^-_il^+_j$
$( l = e, \mu,\tau)$ in the SM is mainly given by
\begin{eqnarray}
\mathcal{H}_{\mathrm{eff}}^{SM} &=& -\frac{4 G_F}{\sqrt{2}} \, V_{ts}^*
V_{tb} \, \Bigl\{ \sum_{i=1}^{6} {C}_i (\mu) \mathcal{O}_i (\mu) + C_7 \,%
\frac{e}{16 \pi^2}\, [\bar{s} \sigma_{\mu\nu} (m_s P_L + m_b P_R) b] \,
F^{\mu \nu}  \notag \\
&+& C_9 \,\frac{\alpha_{em}}{4 \pi}\, (\bar{s} \gamma^\mu P_L b) \,
L^\mu_{ij} + C_{10} \,\frac{\alpha_{em}}{4 \pi}\, (\bar{s} \gamma^\mu P_L b)
\, \L ^{5\mu}_{ij} \, \Bigr\},  \label{HSM}
\end{eqnarray}
where $L^\mu_{ij}= \bar{l}_i \gamma_\mu l_j $,~$L^{5\mu}_{ij}= \bar{l}_i
\gamma_\mu \gamma_5 l_j $, and $P_{L,R} = (1 \mp \gamma_5)/2$. The operators
$\mathcal{O}_i$ ($i=1,..6$) correspond to the $P_i$ in ~\cite{Bobeth:1999mk}%
, and $m_b = m_b(\mu)$ is the running $b$-quark mass in the $\overline{%
\mathrm{MS}}$ scheme. We use the SM Wilson coefficients as given in \cite%
{Altmannshofer:2008dz}.

The total effective Hamiltonian for $b \to sl^-_il^+_j$ in the presence of
new physics operators with all the possible Lorentz structure excluding
tensor once can be expressed as \cite{Alok:2010zd}
\begin{equation}
\mathcal{H}_{\mathrm{eff}}(b \to sl^-_il^+_j) = \mathcal{H}_{\mathrm{eff}%
}^{SM} + \mathcal{H}_{\mathrm{eff}}^{VA} + \mathcal{H}_{\mathrm{eff}}^{SP}~,
\label{NP:effHam}
\end{equation}
where $\mathcal{H}_{\mathrm{eff}}^{SM}$ is given by Eq.~(\ref{HSM}), and the
NP contributions are
\begin{eqnarray}
\mathcal{H}_{\mathrm{eff}}^{VA} &=& - N_F \, \Bigl\{ R_V \, (\bar{s}
\gamma^\mu P_L b) \, L^\mu_{ij} + R_A \, (\bar{s} \gamma^\mu P_L b) \,
L^{5\mu}_{ij} + R^{\prime }_V \, (\bar{s} \gamma^\mu P_R b) \, L^\mu_{ij} +
R^{\prime }_A \, (\bar{s} \gamma^\mu P_R b) \, L^{5\mu}_{ij} \Bigr\} ~, \label{VA}\\
\mathcal{H}_{\mathrm{eff}}^{SP} &=& -N_F \, \Bigl\{R_S ~(\bar{s}
P_R b) ~ L_{ij} + R_P ~(\bar{s} P_R b) ~ L^{5}_{ij} + R^{\prime
}_S ~(\bar{s} P_L b) ~L_{ij} + R^{\prime }_P ~(\bar{s} P_L b) ~
L^{5}_{ij} \Bigr\} \label{SP}\;,
\end{eqnarray}%
where $N_F = \frac{4 G_F}{\sqrt{2}} \,\frac{\alpha_{em}}{4\pi} \, V_{ts}^*
V_{tb}$, $L_{ij} = \bar{l}_i l_j$, and $L^{5}_{ij} = \bar{l}_i \gamma_5 l_j$%
. In the above expressions, $R_i$, $R^\prime_i (i = V, A, S, P)$ are the NP
effective couplings which are in general complex.

In terms of all the Wilson coefficients, the branching ratio for the decay $%
B_s \to l^-_i l^+_j$ can be obtained from
\begin{eqnarray}  \label{BrBlilj}
BR(B_s \to l^-_i l^+_j) &=&
\frac{\tau_{B_S}}{\hbar}\frac{f_{B_s}^2 G_F^2
m_{B_s} \alpha_{em}^2 | V_{tb} V^*_{ts}|^2}{64 \pi^3} \sqrt{\Big[1 - \Big(%
\frac{m_i+ m_j}{m_{B_s}}\Big)^2\Big]\Big[1 - \Big(\frac{m_i- m_j}{m_{B_s}}%
\Big)^2\Big]}  \notag \\
&&\Big\{\Big[1 - \Big(\frac{m_i+ m_j}{m_{B_s}}\Big)^2\Big] | F_V (m_i - m_j)
+ F_S |^2 + \Big[1 - \Big(\frac{m_i- m_j}{m_{B_s}}\Big)^2\Big] | F_A (m_i +
m_j) + F_P |^2 \Big\},~~
\end{eqnarray}
where
\begin{eqnarray}  \label{Eqcoup}
F_V &=& C_9 + R_V -R^\prime_V \,,  \notag \\
F_A &=& C_{10} + R_A -R^\prime_A \,,  \notag \\
F_S &=& r_\chi (R_S - R^\prime_S)\,,  \notag \\
F_P &=& r_\chi (R_P - R^\prime_P)\,,
\end{eqnarray}
with $r_\chi = \frac{ m^2_{B_s}}{m_b(\mu) + m_s(\mu)} $. The Wilson
coefficient $F_V$ does not contribute in the lepton flavor conserving decays
$B_s \to \mu^- \mu^+$ and $B_s \to \tau^- \tau^+$. Within the SM, $F_A =
C_{10}$ only contributes to $B_s \to \mu^- \mu^+$ and $B_s \to \tau^- \tau^+$%
. The lepton flavor violating decay $B_s \to \tau \mu$ is not allowed in the
SM. Thus, the Wilslon coefficients $C_9$ and $C_{10}$ do not contribute to
this decay. Next, we look at two models of NP that contribute to $B_{s, d}
\to l^-_i l^+_j$ decays.

\section{Flavor Violating Higgs Decays \label{sec3}}

In this section we consider the model where a neutral scalar boson $X$ with
mass $\sim 125 $ GeV and flavor-violating couplings \cite%
{Blankenburg:2012ex, Harnik:2012pb} mediates the $B_{s,d} \to l^-_i l^+_j$
decays . In this model, the effective Lagrangian which describes the
possible flavor-violating couplings of $X$ to SM fermion pairs in the mass
basis is, \cite{Blankenburg:2012ex, Harnik:2012pb} :
\begin{eqnarray}  \label{125HLag}
{\mathcal{L}}_Y = -m_i \bar{f}^i_L f_R^i - Y_{ij} (\bar{f}^i_L f_R^j) X +
h.c+.....\,,
\end{eqnarray}
where ellipses denote nonrenormalizable couplings involving more than on
Higgs field operator. $f_L = q_L, l_L$ are $\mathrm{SU(2)}_\mathrm{L}$
doublets, $f_R = u_R d_R, \nu_R , l_R$ are the weak singlets. The indices
run over generations and fermion flavors with summation implicitly
understood. In the SM the Higgs coupling are diagonal, $Y_{ij} = (m_i/v)
\delta_{ij}$ with $v = 246 $ GeV, but in general NP models the structure of
the $Y_{ij}$ can be different. The couplings in Eq. (\ref{125HLag}) will
also lead to the decay $b \to sl^-_il^+_j$ at tree level, mediated by a
virtual $X$. The lagrangian gives
\begin{eqnarray}  \label{125HLagRSP}
R_{S(P)} &=& \frac{1}{N_F} \frac{Y_{sb} (Y_{\mu \tau} + Y^{\ast}_{\tau \mu
}) }{4 m_X^2} \,,  \notag \\
R^\prime_{S(P)} &=& \frac{1}{N_F} \frac{Y^{\ast}_{bs} (Y_{\mu \tau} -
Y^{\ast}_{\tau \mu }) }{4 m_X^2} \,,  \notag \\
R_{V(A)} &=& R^\prime_{V(A)}=0.
\end{eqnarray}

The branching ratio for the $B_s \to \mu^- \tau^+$ decay can be obtained
from Eq. (\ref{BrBlilj}) as,
\begin{eqnarray}  \label{125HBrBsmutau}
BR(B_s \to \mu^- \tau^+) &=&
\frac{\tau_{B_S}}{\hbar}\frac{f_{B_s}^2
m^5_{B_s} }{512 \pi (m_b(\mu) + m_s(\mu))^2} \times \sqrt{\Big[1 - \Big(%
\frac{m_\tau+ m_\mu}{m_{B_s}}\Big)^2\Big]\Big[1 - \Big(\frac{m_\tau- m_\mu}{%
m_{B_s}}\Big)^2\Big]}  \notag \\
&\times& \frac{|Y_{sb}-Y^{\ast}_{bs}|^2}{m^4_X} \Big\{\Big[1 - \Big(\frac{%
m_\tau + m_\mu}{m_{B_s}}\Big)^2\Big] | Y_{\mu \tau} + Y^{\ast}_{\tau \mu }
|^2 + \Big[1 - \Big(\frac{m_\tau- m_\mu}{m_{B_s}}\Big)^2\Big] | Y_{\mu \tau}
- Y^{\ast}_{\tau \mu }|^2 \Big\}.~~
\end{eqnarray}
Also, the branching ratio for the $B_s \to l^- l^+ ( l = \mu, \tau)$ decays
is,

\begin{eqnarray}  \label{125HBrBrBlmlp}
BR(B_s \to l^- l^+) &=& \frac{\tau_{B_S}}{\hbar}\frac{f_{B_s}^2
G_F^2
m_{B_s} \alpha_{em}^2 | V_{tb} V^*_{ts}|^2}{64 \pi^3} \sqrt{1 - \frac{4 m^2_l%
}{m^2_{B_s}}}\Big[\Big(1-\frac{4 m^2_l}{m^2_{B_s}}\Big)\Big| \frac{
Re[Y_{ll}](Y_{sb}-Y^{\ast}_{bs})}{2 m^2_X N_F (m_b(\mu) + m_s(\mu)) }\Big|^2
\notag \\
&& + \Big(\frac{2 m_l}{m_{B_s}}\Big)^2 |C_{10}|^2 \Big]\,,
\end{eqnarray}

where we assume $Y_{ll}$ is real and set it to its SM value $Y_{ll} = m_{l}/v
$.

Flavor violating Higgs coupling in Eq. (\ref{125HLag}) can generate flavor
changing neutral currents at tree level. The couplings $Y_{sb}$ and $Y_{bs}$
are constrained by the mass difference $\Delta M_s$ in the $\mathrm{%
B_s\!-\!\,\overline{B}{}_s\,} $ mixing. The $\Delta B=2$ weak Hamiltonian
for this process can be found in the Ref. \cite{Harnik:2012pb}. The
theoretical expression for $\Delta M_s$ is given in Ref.\cite{Datta:2010yq}.
It is found that one can reproduce the measured value of $\Delta M_s =
(17.719 \pm 0.043) ps^{-1} $ \cite{Amhis:2012bh} if  the upper bound on the flavor changing couplings is
\begin{eqnarray}
|Y_{sb}-Y^*_{bs}|\sim 10^{-3}\,.
\end{eqnarray}
We perform a fit to constrain the  leptonic couplings $Y_{\mu \tau}$ and $Y_{\tau \mu}$. The bound on these  couplings are obtained from
the decay rates of $\tau \to \mu \gamma$, $\tau \to 3\mu$, and the magnetic ($%
\delta a_\mu )$ and electric dipole ($d_\mu$) moments. The theoretical
expressions for $\Gamma(\tau \to \mu \gamma)$, $\Gamma(\tau \to 3\mu)$, $%
\Delta a_\mu \equiv a^{exp}_\mu - a^{SM}_\mu$, and $d_\mu$ are taken from
\cite{ Harnik:2012pb}. The experimental bounds for these observables are taken from \cite{Beringer:1900zz, KNak:2010,  Bennett:2006fi} :

\begin{eqnarray}  \label{tauexp}
BR(\tau \to \mu \gamma) &<& 4.4 \times 10^{-8} \nl BR(\tau \to
3\mu ) &<& 2.1 \times 10^{-8} \nl \Delta a_\mu &=& (2.87 \pm 0.63
\pm 0.49)10^{-9} \nl -10 \times 10^{-20} e\,cm &<& d_\mu < 8
\times 10^{-20} e\,cm\,.
\end{eqnarray}

In the fit, $BR(\tau \to \mu \gamma)$, $BR(\tau \to 3\mu )$, $\Delta a_\mu$ and $ d_\mu$ are  constrained by their experimental results in Eq.~(\ref{tauexp}) and $\Delta a_\mu$ is varied within $1 \sigma$ errors. As shown in Fig. 1 (right panel), $\Delta a_\mu$ provides a strong bound on  the couplings $Y_{\mu \tau}$ and $Y_{\tau \mu}$, and in particular requires $|Y_{\mu \tau}+Y^{*}_{\tau \mu}| \gsim 0.09 $ . It is found that one can satisfy the above four experimental bounds if $%
|Y_{\mu \tau}| < 0.064 \ $, $|Y_{\tau \mu }| < 0.061\ $ for $m_X =
125$ GeV. The fit results allow us to estimate the decay rate
$BR(B_s \to \mu \tau) =
BR(B_s \to \mu^- \tau^+) + BR(B_s \to \tau^- \mu^+)$. The variations of the decay rates $BR(B_s \to \mu \tau)$ with $%
|Y_{sb}-Y^{*}_{bs}|$ and $|Y_{\mu \tau}+Y^{*}_{\tau \mu }|$ are
shown in Fig. 1. Our result predict that the $BR(B_s \to \mu \tau)
$ can be as large as $\sim
6 \times 10^{-6}$ for $m_X = 125$ GeV and $f_{B_s} = 0.229 \pm 0.006 $ GeV \cite%
{Charles:2004jd}.


\begin{figure}[htb]
\begin{center}
$%
\begin{array}{cc}
\epsfig{file=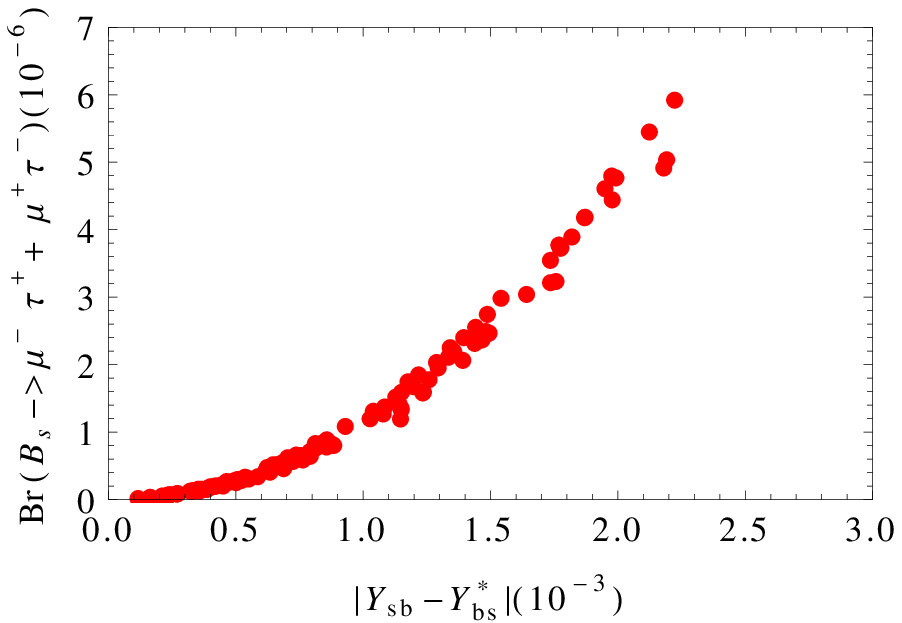,height=6.0cm,width=6.5cm,angle=0}~~~~~~~ %
\epsfig{file=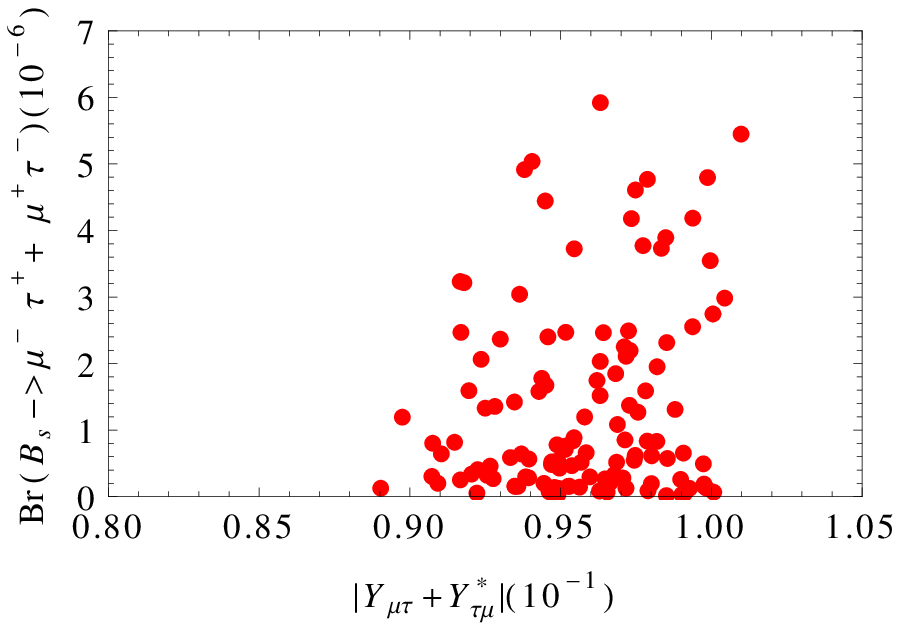,height=6.0cm,width=6.5cm,angle=0} \vspace{%
-0.8cm} &
\end{array}%
$%
\end{center}
\caption{ The variation of \emph{$BR(B_s \to \protect\mu^- \protect\tau%
^+~~+~ \protect\mu^+ \protect\tau^-)$ with the couplings $|Y_{sb}-Y^{*}_{bs}|
$ and $|Y_{\protect\mu \protect\tau}-Y^{*}_{\protect\tau \protect\mu }|$ for
$m_X = 125$ GeV and $f_{B_s} = .229 \pm 0.006 $ GeV. Scatter points are
allowed by $\Delta M_s^{exp}$ and the experimental bounds in Eq. (\protect\ref%
{tauexp}). }}
\label{hgg}
\end{figure}

In this model, we obtain the branching ratio $BR(B_s \to \mu^-
\mu^+)< 4.3 \times 10^{-9}$ using $Y_{\mu \mu} = m_\mu/v$ with $v
= 24 6$ GeV, $m_X = 125 $ GeV and $f_{B_s} =0.229 \pm 0.006 $ GeV.
The corresponding SM prediction is $\approx 3.17 \times 10^{-9}$.
Our results for $BR(B_s \to \mu^- \mu^+)$ is consistent with the
current upper limit $BR(B_s \to \mu^- \mu^+)\equiv \left[1.1 - 6.4
\right] \times 10^{-9}$ at 95\% C.L. in Ref. \cite{:2012ct}.
Also, we obtain the branching ratio $BR(B_s \to \tau^- \tau^+)<
8.1 \times 10^{-7}$ using $Y_{\tau \tau} = m_\tau/v$ and $m_X =
125$ GeV. The values of this branching ratio in the SM is $\approx
6.81 \times 10^{-7}$. The latest LHCb-measurement of $\Gamma_d/
\Gamma_s$ implies a limit of $BR(B_s \to \tau^- \tau^+)< 3 \%$
\cite{Bediaga:2012py}. In passing we note that, the large width
difference in the $B_s$ meson system can changes these results by
$\sim {\cal{O}}(10\%) $ as pointed out in
\cite{DeBruyn:2012wj,DeBruyn:2012wk, Datta:2012ky}.

Finally, in this model we obtain the tau longitudinal polarization fraction
for the decays $B_s \to \mu^- \tau^+$:
\begin{eqnarray}  \label{taupol}
P_L & = & \frac{Br_{B_s \to \mu^- \tau^+}[\lambda_\tau = 1/2]-Br_{B_s \to
\mu^- \tau^+}[\lambda_\tau = -1/2]}{Br_{B \to \mu^- \tau^+}[\lambda_\tau =
1/2]+Br_{B_s \to \mu^- \tau^+}[\lambda_\tau = -1/2]} = \sqrt{\Big[1 - \Big(%
\frac{m_\tau+ m_\mu}{m_{B_s}}\Big)^2\Big]\Big[1 - \Big(\frac{m_\tau- m_\mu}{%
m_{B_s}}\Big)^2\Big]}\times  \notag \\
&& \Big(2 (| Y_{\mu \tau} |^2 -| Y^*_{\tau \mu} |^2)\Big)/ \Big\{\Big[1 - %
\Big(\frac{m_\tau + m_\mu}{m_{B_s}}\Big)^2\Big] | Y_{\mu \tau} +
Y^{\ast}_{\tau \mu } |^2 + \Big[1 - \Big(\frac{m_\tau- m_\mu}{m_{B_s}}\Big)^2%
\Big] | Y_{\mu \tau} - Y^{\ast}_{\tau \mu }|^2 \Big\}\,.
\end{eqnarray}
The result indicates $P_L$ does not depend on the $b\to s$ couplings. Fig.~%
\ref{hgg} shows the dependency of $P_L$ on the quantity $(| Y_{\mu \tau} |^2
-| Y^*_{\tau \mu} |^2)$.

\begin{figure}[htb]
\begin{center}
$%
\begin{array}{cc}
\epsfig{file=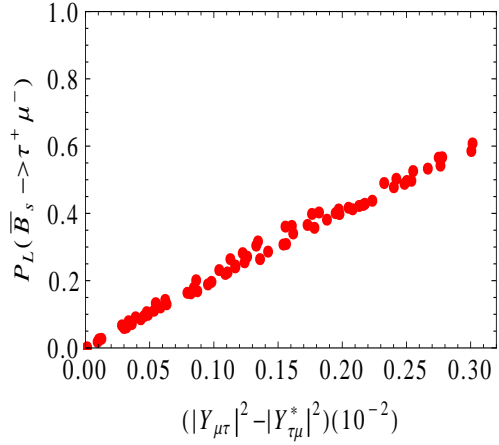,height=6.0cm,width=6.5cm,angle=0} \vspace{%
-0.8cm} &
\end{array}%
$%
\end{center}
\caption{Figure shows the $(| Y_{\protect\mu \protect\tau} |^2 -| Y^*_{%
\protect\tau \protect\mu} |^2)$ dependence of \emph{$P_L(B_s \to \protect\mu%
^- \protect\tau^+)$. Scatter points are allowed by $\Delta M_s^{exp}$ and
the experimental bounds in Eq. (\protect\ref{tauexp}). }}
\label{hgg}
\end{figure}

The couplings $Y_{db}$ and $Y_{bd}$ for the decay $BR(B \to \mu
\tau) = BR(B \to \mu^- \tau^+) + BR(B \to \tau^- \mu^+)$ are
similarly constrained by the mass difference $\Delta M_d =(0.507
\pm 0.004) ps^{-1} $ \cite{Lenz:2012mb}. The measured $\Delta M_d$
can be reproduced if  the upper bound on the flavor changing
couplings is $|Y_{db}-Y^*_{bd}|\sim 10^{-4} $. Using this result,
we findd the decay rate $BR(B \to \mu \tau) < 2 \times 10^{-7}$.

\section{Supersymmetry \label{sec4}}

The supersymmetry contributions to the process $B_{s}\rightarrow \mu
^{+}\tau ^{-}$ are given by the one-loop box diagrams, shown in Fig \ref%
{loop}, where charginos and neutralinos are exchanged. The interaction of
charginos, neutralinos can be written as
\begin{figure}[htb]
\begin{center}
$%
\begin{array}{cc}
\epsfig{file=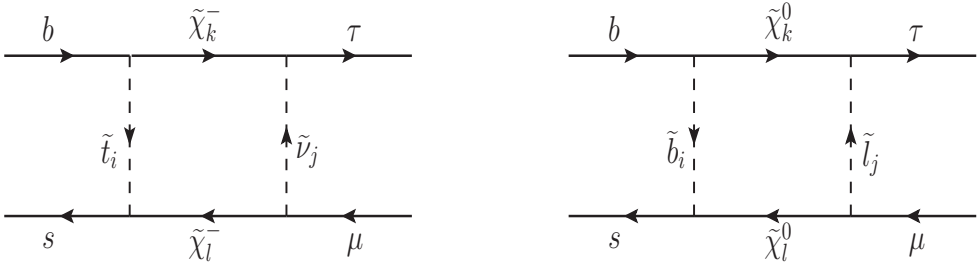,height=3.5cm,width=13cm,angle=0}\vspace{-0.4cm}
&
\end{array}
$%
\end{center}
\caption{One loop SUSY contributions to $B_s \to \protect\tau^- \protect\mu^+
$ . }
\label{loop}
\end{figure}
\begin{eqnarray}
\mathcal{L} &=&\sum_{j=1}^{2}\overline{\tilde{\chi}_{j}^{-}}\Big[\tilde{u}%
_{i}^{\dag }\left( G_{IJij}^{U_{L}}P_{L}+G_{IJij}^{U_{R}}P_{R}\right) d^{I}+%
\tilde{\nu}_{J}^{\dag }\left(
G_{IJj}^{L_{L}}P_{L}+G_{IJj}^{L_{R}}P_{R}\right) l^{I}\Big]+hc  \notag \\
&+&\sum_{k=1}^{4}\overline{\tilde{\chi}_{k}^{0}}\Big[\tilde{d}_{i}^{\dag
}\left( Z_{Iik}^{D_{L}}P_{L}+Z_{Iik}^{D_{R}}P_{R}\right) d^{I}+\tilde{l}%
_{i}^{\dag }\left( Z_{Iik}^{L_{L}}P_{L}+Z_{Iik}^{L_{R}}P_{R}\right) l^{I}%
\Big]+hc,
\end{eqnarray}%
where the mixing matrices in the super-CKM basis are given by \cite%
{Kruger:2000ff}%
\begin{eqnarray}
G_{IJij}^{U_{L}} &=&g\Big[-V_{j1}^{\ast }(\Gamma ^{U_{L}})_{Ji}+V_{j2}^{\ast
}(\Gamma ^{U_{R}})_{Ji}\frac{m_{u^{J}}}{\sqrt{2}M_{W}\sin \beta }\Big]%
V_{JI}^{_{^{CKM}}},\text{ \ }G_{IJij}^{U_{R}}=gU_{j2}(\Gamma ^{U_{L}})_{Ji}%
\frac{m_{d^{I}}}{\sqrt{2}m_{W}\cos \beta }V_{JI}^{_{CKM}},~~ \\
G_{IJj}^{L_{L}} &=&-gV_{j1}^{\ast }(\Gamma ^{\tilde{\nu}})_{IJ},\text{ \ \ \
\ \ \ \ \ \ \ \ \ \ \ \ \ \ \ \ \ \ \ \ \ \ \ \ \ \ \ \ \ \ \ \ \ \ \ \ \ \
\ \ \ \ \ \ \ \ \ \ }G_{IJj}^{L_{R}}=gU_{j2}(\Gamma ^{\tilde{\nu}})_{IJ}%
\frac{m_{l^{I}}}{\sqrt{2}m_{W}\cos \beta }, \\
Z_{Iik}^{L_{L}} &=&-\frac{g}{\sqrt{2}}\Big[(N_{k2}^{\ast }+\tan \theta
_{W}N_{k1}^{\ast })(\Gamma ^{L_{L}})_{Ii}-N_{k3}^{\ast }(\Gamma
^{L_{R}})_{Ii}\frac{m_{l}}{m_{W}\cos \beta }\Big], \\
Z_{Iik}^{L_{R}} &=&-\frac{g}{\sqrt{2}}\Big[2\tan \theta _{W}N_{k1}(\Gamma
^{L_{R}})_{Ii}+N_{k3}(\Gamma ^{L_{L}})_{Ii}\frac{m_{l}}{m_{W}\cos \beta }%
\Big], \\
Z_{Iik}^{D_{L}} &=&-\frac{g}{\sqrt{2}}\Big[(N_{k2}^{\ast }+\frac{1}{2}\tan
\theta _{W}N_{k1}^{\ast })(\Gamma ^{D_{L}})_{Ii}+N_{k3}^{\ast }(\Gamma
^{D_{R}})_{Ii}\frac{m_{d}}{M_{W}\cos \beta }\Big], \\
Z_{Iik}^{D_{R}} &=&-\frac{g}{\sqrt{2}}\Big[\frac{2}{3}\tan \theta
_{W}N_{k1}(\Gamma ^{D_{R}})_{Ii}+N_{k3}(\Gamma ^{D_{L}})_{Ii}\frac{m_{d}}{%
m_{W}\cos \beta }\Big],
\end{eqnarray}%
where $I,J=1..3$ and $i=1..6$. The matrices $\Gamma $'s are the matrices
that diagonalize the squark and slepton mass matrices, while $U$ and $V$ are
the unitary matrices that diagonalize the charged mass matrices. Finally, $N$
is the matrix that diagonalizes the neutralino mass matrix.

\subsection{Chargino Contribution}

The general expression for the amplitude of chargino contribution
to $b\to s l^+ l^-$ is given by%
\begin{eqnarray}
A_{\widetilde{\chi }^{\pm }} &=&-\frac{i}{16\pi
^{2}}\Big(m_{\widetilde{\chi
}_{l}^{+}}m_{\widetilde{\chi }_{k}^{+}}B(m_{\widetilde{\chi }_{k}^{+}},m_{%
\widetilde{\chi }_{l}^{+}},m_{\widetilde{t}_{i}},m_{\widetilde{\nu }_{j}})%
\Big[\frac{1}{2}G_{2Nl}^{L_{L}}G_{3Nk}^{L_{R}\ast
}G_{3Hik}^{U_{L}}G_{2Jil}^{\ast U_{R}}(\overline{s}P_{L}b)(\overline{\tau }%
P_{L}\mu ) \nonumber\\
&&+\frac{1}{2}G_{2Nl}^{L_{R}}G_{3Nk}^{L_{L}\ast
}G_{3Hik}^{U_{R}}G_{2Jil}^{\ast U_{L}}(\overline{s}P_{R}b)(\overline{\tau }%
P_{R}\mu )+\frac{1}{4}G_{3Nk}^{L_{L}\ast
}G_{2Nl}^{L_{L}}G_{3Hik}^{U_{R}}G_{2Jil}^{\ast
U_{R}}(\overline{s}\gamma
^{\mu }P_{L}b)(\overline{\tau }\gamma _{\mu }P_{R}\mu ) \nonumber\\
&&+\frac{1}{4}G_{3Nk}^{L_{R}\ast
}G_{2Nl}^{L_{R}}G_{3Hik}^{U_{L}}G_{2Jil}^{\ast
U_{L}}(\overline{s}\gamma
^{\mu }P_{R}b)(\overline{\tau }\gamma _{\mu }P_{L}\mu )\Big] \nonumber\\
&&+\frac{1}{4}S(m_{\widetilde{\chi }_{k}^{+}},m_{\widetilde{\chi }%
_{l}^{+}},m_{\widetilde{t}_{i}},m_{\widetilde{\nu }_{j}})\Big[%
G_{2Nl}^{L_{L}}G_{3Nk}^{L_{R}\ast }G_{3Hik}^{U_{L}}G_{2Jil}^{\ast U_{R}}(%
\overline{s}\gamma ^{\mu }P_{L}b)(\overline{\tau }\gamma _{\mu }P_{L}\mu ) \nonumber\\
&&+G_{2Nl}^{L_{R}}G_{3Nk}^{L_{L}\ast }G_{3Hik}^{U_{R}}G_{2Jil}^{\ast U_{L}}(%
\overline{s}\gamma ^{\mu }P_{R}b)(\overline{\tau }\gamma _{\mu }P_{R}\mu ) \nonumber\\
&&+2G_{3Nk}^{L_{L}\ast }G_{2Nl}^{L_{L}}G_{3Hik}^{U_{R}}G_{2Jil}^{\ast U_{R}}(%
\overline{s}P_{L}b)(\overline{\tau }P_{R}\mu ) \nonumber\\
&&+2G_{3Nk}^{L_{R}\ast }G_{2Nl}^{L_{R}}G_{3Hik}^{U_{L}}G_{2Jil}^{\ast U_{L}}(%
\overline{s}P_{R}b)(\overline{\tau }P_{L}\mu )\Big]\Big).
\end{eqnarray}
Therefore, one can write this amplitude as %
\be
A_{\widetilde{\chi }^{\pm }}=A_{\widetilde{\chi }^{\pm }}^{S-P}+A_{%
\widetilde{\chi }^{\pm }}^{V-A},%
\ee%
with
\begin{eqnarray}
A_{\widetilde{\chi }^{\pm }}^{S-P} &=&-\frac{i}{16\pi ^{2}}\Big(%
(C_{SRR}+C_{SRL})(\overline{s}P_{R}b)\overline{\tau }\mu +(C_{SLL}+C_{SLR})(%
\overline{s}P_{L}b)\overline{\tau }\mu \nonumber\\
&&+(C_{SRR}-C_{SRL})(\overline{s}P_{R}b)\overline{\tau }\gamma
^{5}\mu +(C_{SLR}-C_{SLL})(\overline{s}P_{L}b)\overline{\tau }\gamma
^{5}\mu \Big),
\end{eqnarray}
and
\begin{eqnarray}
A_{\widetilde{\chi }^{\pm }}^{V-A} &=&-\frac{i}{16\pi ^{2}}\Big(%
(C_{VLL}+C_{VLR})(\overline{s}\gamma ^{\mu }P_{L}b)(\overline{\tau
}\gamma
_{\mu }\mu )+(C_{VRR}+C_{VRL})(\overline{s}\gamma ^{\mu }P_{R}b)(\overline{%
\tau }\gamma _{\mu }\mu ) \nonumber\\
&&+(C_{VLR}-C_{VLL}+)(\overline{s}\gamma ^{\mu }P_{L}b)(\overline{\tau }%
\gamma _{\mu }\gamma ^{5}\mu )+(C_{VRR}-C_{VRL})(\overline{s}\gamma
^{\mu }P_{R}b)(\overline{\tau }\gamma _{\mu }\gamma ^{5}\mu )\Big)
\end{eqnarray}
where
\begin{eqnarray}
C_{SLL} &=&\frac{g^{4}}{2m_{W}^{2}}V_{tb}^{CKM}V_{ts}^{CKM\ast
}V_{l1}^{\ast
}U_{l2}^{\ast }\Gamma _{33}^{\widetilde{\nu }\ast }\Gamma _{23}^{\widetilde{%
\nu }}\Big[U_{k2}^{\ast }V_{k1}^{\ast }\Gamma _{3i}^{U_{L}}\Gamma
_{3i}^{U_{L}\ast }-\frac{m_{t}}{\sqrt{2}m_{W}\sin \beta
}U_{k2}^{\ast
}V_{k2}^{\ast }\Gamma _{3i}^{U_{R}}\Gamma _{3i}^{U_{L}\ast }\Big] \nonumber\\
&&\times \frac{m_{s}m_{\tau }m_{\widetilde{\chi
}_{l}^{+}}m_{\widetilde{\chi
}_{k}^{+}}}{4\cos ^{2}\beta }B(m_{\widetilde{\chi }_{k}^{+}},m_{\widetilde{%
\chi }_{l}^{+}},m_{\widetilde{t}_{i}},m_{\widetilde{\nu }_{j}}) \\
C_{SRR} &=&\frac{g^{4}}{2m_{W}^{2}}V_{tb}^{CKM}V_{ts}^{CKM\ast
}V_{k1}U_{k2}\Gamma _{33}^{\widetilde{\nu }\ast }\Gamma _{23}^{\widetilde{%
\nu }}\Big[U_{l2}V_{l1}\Gamma _{3i}^{U_{L}}\Gamma _{3i}^{U_{L}\ast }-\frac{%
m_{t}}{\sqrt{2}m_{W}\sin \beta }U_{l2}V_{l2}\Gamma _{3i}^{U_{R}\ast
}\Gamma
_{3i}^{U_{L}}\Big] \nonumber\\
&&\times \frac{m_{b}m_{\mu }m_{\widetilde{\chi }_{l}^{+}}m_{\widetilde{\chi }%
_{k}^{+}}}{4\cos ^{2}\beta }B(m_{\widetilde{\chi }_{k}^{+}},m_{\widetilde{%
\chi }_{l}^{+}},m_{\widetilde{t}_{i}},m_{\widetilde{\nu }_{j}}) \\
C_{SLR} &=&\frac{g^{4}}{2m_{W}^{2}}V_{tb}^{CKM}V_{ts}^{CKM\ast
}\Gamma _{33}^{\widetilde{\nu }\ast }\Gamma _{23}^{\widetilde{\nu
}}(U_{l2}^{\ast
}V_{l1}^{\ast })(U_{k2}V_{k1})\Gamma _{3i}^{U_{L}}\Gamma _{3i}^{U_{L}\ast }%
\frac{m_{b}m_{s}}{4\cos ^{2}\beta }S(m_{\widetilde{\chi }_{k}^{+}},m_{%
\widetilde{\chi }_{l}^{+}},m_{\widetilde{t}_{i}},m_{\widetilde{\nu }_{j}})\\
C_{SRL} &=&\frac{g^{4}}{2m_{W}^{2}}V_{tb}^{CKM}V_{ts}^{CKM\ast
}\Gamma
_{33}^{\widetilde{\nu }\ast }\Gamma _{23}^{\widetilde{\nu }}\Big[%
(U_{k2}^{\ast }V_{k1}^{\ast })(U_{l2}^{\ast }V_{l1})\Gamma
_{3i}^{U_{L}}\Gamma _{3i}^{U_{L}\ast }+(U_{k2}^{\ast }V_{k2}^{\ast
})(U_{l2}^{\ast }V_{l2})\Gamma _{3i}^{U_{R}}\Gamma _{3i}^{U_{R}\ast }\frac{%
m_{t}^{2}}{2m_{W}^{2}\sin ^{2}\beta } \nonumber\\
&&-\frac{m_{t}}{\sqrt{2}m_{W}\sin \beta }\Big((U_{k2}^{\ast
}V_{k1}^{\ast })(U_{l2}^{\ast }V_{l2})\Gamma _{3i}^{U_{L}}\Gamma
_{3i}^{U_{R}\ast }+(U_{k2}^{\ast }V_{k2}^{\ast })(U_{l2}^{\ast
}V_{l1})\Gamma
_{3i}^{U_{L}\ast }\Gamma _{3i}^{U_{R}}\Big)\Big] \nonumber\\
&&\times \frac{m_{\mu }m_{\tau }}{4\cos ^{2}\beta }S(m_{\widetilde{\chi }%
_{k}^{+}},m_{\widetilde{\chi }_{l}^{+}},m_{\widetilde{t}_{i}},m_{\widetilde{%
\nu }_{j}})
\end{eqnarray}
and
\begin{eqnarray}
C_{VLL} &=&\frac{g^{4}}{2m_{W}^{2}}V_{tb}^{CKM}V_{ts}^{CKM\ast
}V_{l1}^{\ast
}U_{l2}^{\ast }\Gamma _{33}^{\widetilde{\nu }\ast }\Gamma _{23}^{\widetilde{%
\nu }}\Big(U_{k2}^{\ast }V_{k1}^{\ast }\Gamma _{3i}^{U_{L}}\Gamma
_{3i}^{U_{L}\ast }-\frac{m_{t}}{\sqrt{2}m_{W}\sin \beta
}U_{k2}^{\ast
}V_{k2}^{\ast }\Gamma _{3i}^{U_{R}}\Gamma _{3i}^{U_{L}\ast }\Big) \nonumber\\
&&\times \frac{m_{s}m_{\tau }}{8\cos ^{2}\beta }S(m_{\widetilde{\chi }%
_{k}^{+}},m_{\widetilde{\chi }_{l}^{+}},m_{\widetilde{t}_{i}},m_{\widetilde{%
\nu }_{j}}) \\
C_{VRR} &=&\frac{g^{4}}{2m_{W}^{2}}V_{tb}^{CKM}V_{ts}^{CKM\ast
}V_{k1}U_{k2}\Gamma _{33}^{\widetilde{\nu }\ast }\Gamma _{23}^{\widetilde{%
\nu }}\Big(U_{l2}V_{l1}\Gamma _{3i}^{U_{L}}\Gamma _{3i}^{U_{L}\ast }-\frac{%
m_{t}}{\sqrt{2}m_{W}\sin \beta }U_{l2}V_{l2}\Gamma _{3i}^{U_{R}\ast
}\Gamma
_{3i}^{U_{L}}\Big) \nonumber\\
&&\times \frac{m_{b}m_{\mu }}{8\cos ^{2}\beta }S(m_{\widetilde{\chi }%
_{k}^{+}},m_{\widetilde{\chi }_{l}^{+}},m_{\widetilde{t}_{i}},m_{\widetilde{%
\nu }_{j}}) \\
C_{VLR} &=&\frac{g^{4}}{2m_{W}^{2}}V_{tb}^{CKM}V_{ts}^{CKM\ast
}\Gamma _{33}^{\widetilde{\nu }\ast }\Gamma _{23}^{\widetilde{\nu
}}(U_{l2}^{\ast
}V_{l1}^{\ast })(U_{k2}V_{k1})\Gamma _{3i}^{U_{L}}\Gamma _{3i}^{U_{L}\ast }%
\frac{m_{b}m_{s}m_{\widetilde{\chi }_{l}^{+}}m_{\widetilde{\chi }_{k}^{+}}}{%
8\cos ^{2}\beta }B(m_{\widetilde{\chi }_{k}^{+}},m_{\widetilde{\chi }%
_{l}^{+}},m_{\widetilde{t}_{i}},m_{\widetilde{\nu }_{j}}) \\
C_{VLR} &=&\frac{g^{4}}{2m_{W}^{2}}V_{tb}^{CKM}V_{ts}^{CKM\ast
}\Gamma _{33}^{\widetilde{\nu }\ast }\Gamma _{23}^{\widetilde{\nu
}}(U_{l2}^{\ast
}V_{l1}^{\ast })(U_{k2}V_{k1})\Gamma _{3i}^{U_{L}}\Gamma _{3i}^{U_{L}\ast }%
\frac{m_{b}m_{s}m_{\widetilde{\chi }_{l}^{+}}m_{\widetilde{\chi }_{k}^{+}}}{%
8\cos ^{2}\beta }B(m_{\widetilde{\chi }_{k}^{+}},m_{\widetilde{\chi }%
_{l}^{+}},m_{\widetilde{t}_{i}},m_{\widetilde{\nu }_{j}}) \\
C_{VRL} &=&\frac{g^{4}}{2m_{W}^{2}}V_{tb}^{CKM}V_{ts}^{CKM\ast
}\Gamma
_{33}^{\widetilde{\nu }\ast }\Gamma _{23}^{\widetilde{\nu }}\Big[%
(U_{k2}^{\ast }V_{k1}^{\ast })(U_{l2}^{\ast }V_{l1})\Gamma
_{3i}^{U_{L}}\Gamma _{3i}^{U_{L}\ast }+(U_{k2}^{\ast }V_{k2}^{\ast
})(U_{l2}^{\ast }V_{l2})\Gamma _{3i}^{U_{R}}\Gamma _{3i}^{U_{R}\ast }\frac{%
m_{t}^{2}}{2m_{W}^{2}\sin ^{2}\beta } \nonumber\\
&&-\frac{m_{t}}{\sqrt{2}m_{W}\sin \beta }\Big((U_{k2}^{\ast
}V_{k1}^{\ast })(U_{l2}^{\ast }V_{l2})\Gamma _{3i}^{U_{L}}\Gamma
_{3i}^{U_{R}\ast }+(U_{k2}^{\ast }V_{k2}^{\ast })(U_{l2}^{\ast
}V_{l1})\Gamma
_{3i}^{U_{L}\ast }\Gamma _{3i}^{U_{R}}\Big)\Big] \nonumber\\
&&\times \frac{m_{\mu }m_{\tau }m_{\widetilde{\chi }_{l}^{+}}m_{\widetilde{%
\chi }_{k}^{+}}}{8\cos ^{2}\beta }S(m_{\widetilde{\chi }_{k}^{+}},m_{%
\widetilde{\chi }_{l}^{+}},m_{\widetilde{t}_{i}},m_{\widetilde{\nu
}_{j}}).
\end{eqnarray}

Therefore, from Eqs.(\ref{VA}) and (\ref{SP}) one finds%
\bea%
R_{V,A} &=& \frac{i}{16\pi^2 } \frac{1}{N_F} \left(C_{VLR} \pm
C_{VLL}\right),\\
R'_{V,A} &=& \frac{i}{16\pi^2 } \frac{1}{N_F} \left(C_{VRR} \pm
C_{VRL}\right),\\
R_{S,P} &=& \frac{i}{16\pi^2 } \frac{1}{N_F} \left(C_{SRR} \pm
C_{SRL}\right),\\
R'_{S,P} &=& \frac{i}{16\pi^2 } \frac{1}{N_F} \left(C_{SLR} \pm
C_{SLL}\right).%
\eea
These co-efficients can be used in Eq.~\ref{BrBlilj} to calculate the branching ratio for
$B_{s}\rightarrow \mu^{+}\tau ^{-}$.

The loop functions $B(m_{\widetilde{\chi
}_{k}^{+}},m_{\widetilde{\chi }_{l}^{+}},
m_{\widetilde{t}_{i}},m_{\widetilde{\nu }_{j}})$ and
$S(m_{\widetilde{\chi }_{k}^{+}},m_{\widetilde{\chi }_{l}^{+}},
m_{\widetilde{t}_{i}},m_{\widetilde{\nu}_{j}})$ are given by%
\begin{eqnarray}
B(m_{\widetilde{\chi }_{k}^{+}},m_{\widetilde{\chi }_{l}^{+}},m_{\widetilde{t%
}_{i}},m_{\widetilde{\nu }_{j}}) &=&\frac{m_{\widetilde{t}_{i}}^{2}}{(m_{%
\widetilde{t}_{i}}^{2}-m_{\widetilde{\nu }_{j}}^{2})(m_{\widetilde{t}%
_{i}}^{2}-m_{\widetilde{\chi }_{k}^{+}}^{2})(m_{\widetilde{t}_{i}}^{2}-m_{%
\widetilde{\chi }_{l}^{+}}^{2})}\ln \frac{m_{\widetilde{t}_{i}}^{2}}{m_{%
\widetilde{\chi }_{l}^{+}}^{2}} \nonumber\\
&&+\frac{m_{\widetilde{\nu }_{j}}^{2}}{(m_{\widetilde{\nu }_{j}}^{2}-m_{%
\widetilde{t}_{i}}^{2})(m_{\widetilde{\nu }_{j}}^{2}-m_{\widetilde{\chi }%
_{k}^{+}}^{2})(m_{\widetilde{\nu }_{j}}^{2}-m_{\widetilde{\chi }%
_{l}^{+}}^{2})}\ln \frac{m_{\widetilde{\nu }_{j}}^{2}}{m_{\widetilde{\chi }%
_{l}^{+}}^{2}} \nonumber\\
&&+\frac{m_{\widetilde{\chi }_{k}^{+}}^{2}}{(m_{\widetilde{\chi }%
_{k}^{+}}^{2}-m_{\widetilde{t}_{i}}^{2})(m_{\widetilde{\chi }%
_{k}^{+}}^{2}-m_{\widetilde{\nu }_{j}}^{2})(m_{\widetilde{\chi }%
_{k}^{+}}^{2}-m_{\widetilde{\chi }_{l}^{+}}^{2})}\ln \frac{m_{\widetilde{%
\chi }_{k}^{+}}^{2}}{m_{\widetilde{\chi }_{l}^{+}}^{2}},
\end{eqnarray}
and
\begin{eqnarray}
S(m_{\widetilde{\chi }_{k}^{+}},m_{\widetilde{\chi }_{l}^{+}},m_{\widetilde{t%
}_{i}},m_{\widetilde{\nu }_{j}}) &=&\frac{m_{\widetilde{t}_{i}}^{4}}{(m_{%
\widetilde{t}_{i}}^{2}-m_{\widetilde{\nu }_{j}}^{2})(m_{\widetilde{t}%
_{i}}^{2}-m_{\widetilde{\chi }_{k}^{+}}^{2})(m_{\widetilde{t}_{i}}^{2}-m_{%
\widetilde{\chi }_{l}^{+}}^{2})}\ln \frac{m_{\widetilde{t}_{i}}^{2}}{m_{%
\widetilde{\chi }_{l}^{+}}^{2}} \nonumber\\
&&+\frac{m_{\widetilde{\nu }_{j}}^{4}}{(m_{\widetilde{\nu }_{j}}^{2}-m_{%
\widetilde{t}_{i}}^{2})(m_{\widetilde{\nu }_{j}}^{2}-m_{\widetilde{\chi }%
_{k}^{+}}^{2})(m_{\widetilde{\nu }_{j}}^{2}-m_{\widetilde{\chi }%
_{l}^{+}}^{2})}\ln \frac{m_{\widetilde{\nu }_{j}}^{2}}{m_{\widetilde{\chi }%
_{l}^{+}}^{2}} \nonumber\\
&&+\frac{m_{\widetilde{\chi }_{k}^{+}}^{4}}{(m_{\widetilde{\chi }%
_{k}^{+}}^{2}-m_{\widetilde{t}_{i}}^{2})(m_{\widetilde{\chi }%
_{k}^{+}}^{2}-m_{\widetilde{\nu }_{j}}^{2})(m_{\widetilde{\chi }%
_{k}^{+}}^{2}-m_{\widetilde{\chi }_{l}^{+}}^{2})}\ln \frac{m_{\widetilde{%
\chi }_{k}^{+}}^{2}}{m_{\widetilde{\chi }_{l}^{+}}^{2}}.
\end{eqnarray}
where $i,j,k,l=1..2$ denote the heavy and light sparticles
(stop-quark, sneutrino and chargino).  It is worth mentioning that
the above expression for the chargino amplitude is consistent with
the amplitude reported in Ref.\cite{Bertolini:1990if} and also
with the expressions given in the second paper in
Ref.\cite{btaumu_old}. We have expressed it in the above form to
be close to the the new physics amplitudes discussed in
Ref.\cite{Alok:2010zd} and adopted in the previous section.
Finally, in deriving these expression, the following hadronic
matrix elements have been assumed:%
\bea%
\left\langle 0 \left\vert \overline{s}\gamma^{5} b \right\vert \overline{B}_{s}(p)%
\right\rangle & = & -\frac{i}{2}
f_{B_{s}}\frac{m_{B_{s}}^{2}}{m_{b}+m_{s}},\text{ \ \ \ \ \
\ }\left\langle 0\left\vert \overline{s} b \right\vert \overline{B}_{s}(p)%
\right\rangle =0 ,\\
\left\langle 0\left\vert \overline{s}\gamma_\mu \gamma^{5} b \right\vert \overline{B}_{s}(p)%
\right\rangle & = & \frac{i}{2} p_{\mu} f_{B_{s}} ,\text{ \ \ \ \
\ \ }
\left\langle 0\left\vert \overline{s} \gamma_\mu b\right\vert \overline{B}_{s}(p)%
\right\rangle =0 ,%
\eea%
where $p_\mu$ is the momentum of the $B_s$ meson of mass $B_s$.

\subsection{Neutralion Contribution}

Similar to the chargino contribution, we can write the amplitude
of $b \to s \mu^+ \tau^-$ through the neutralino exchange as
follow:
\begin{eqnarray}
A_{\widetilde{\chi }^{0}} &=&-\frac{i}{16\pi ^{2}}\Big(m_{\widetilde{\chi }%
_{l}^{0}}m_{\widetilde{\chi }_{k}^{0}}B(m_{\widetilde{\chi }_{k}^{0}},m_{%
\widetilde{\chi }_{l}^{0}},m_{\widetilde{b}_{i}},m_{\widetilde{l}_{j}})\Big[%
\frac{1}{2}Z_{2jl}^{L_{L}}Z_{3jk}^{L_{R}\ast
}Z_{3ik}^{D_{L}}Z_{2il}^{\ast
D_{R}}(\overline{s}P_{L}b)(\overline{\tau }P_{L}\mu ) \nonumber\\
&&+\frac{1}{2}Z_{3jk}^{L_{L}\ast
}Z_{2jl}^{L_{R}}Z_{3ik}^{D_{R}}Z_{2il}^{\ast D_{L}}(\overline{s}P_{R}b)(%
\overline{\tau }P_{R}\mu )+\frac{1}{4}Z_{2jl}^{L_{L}}Z_{3jk}^{\ast
L_{L}}Z_{3ik}^{D_{R}}Z_{2il}^{\ast D_{R}}(\overline{s}\gamma ^{\mu }P_{L}b)(%
\overline{\tau }\gamma _{\mu }P_{R}\mu ) \nonumber\\
&&+\frac{1}{4}Z_{2jl}^{L_{R}}Z_{3jk}^{\ast
L_{R}}Z_{3ik}^{D_{L}}Z_{2il}^{\ast D_{L}}(\overline{s}\gamma ^{\mu }P_{R}b)(%
\overline{\tau }\gamma _{\mu }P_{L}\mu )\Big] \nonumber\\
&&+\frac{1}{4}S(m_{\widetilde{\chi }_{k}^{+}},m_{\widetilde{\chi }%
_{l}^{+}},m_{\widetilde{b}_{i}},m_{\widetilde{\nu }%
_{j}})\Big[Z_{2jl}^{L_{L}}Z_{3jk}^{L_{R}\ast
}Z_{3ik}^{D_{L}}Z_{2il}^{\ast D_{R}}(\overline{s}\gamma ^{\mu
}P_{L}b)(\overline{\tau }\gamma _{\mu
}P_{L}\mu ) \nonumber\\
&&+Z_{3jk}^{L_{L}\ast }Z_{2jl}^{L_{R}}Z_{3ik}^{D_{R}}Z_{2il}^{\ast D_{L}}(%
\overline{s}\gamma ^{\mu }P_{R}b)(\overline{\tau }\gamma _{\mu }P_{R}\mu ) \nonumber\\
&&+2Z_{2jl}^{L_{L}}Z_{3jk}^{\ast L_{L}}Z_{3ik}^{D_{R}}Z_{2il}^{\ast D_{R}}(%
\overline{s}P_{L}b)(\overline{\tau }P_{R}\mu ) \nonumber\\
&&+2Z_{2jl}^{L_{R}}Z_{3jk}^{\ast L_{R}}Z_{3ik}^{D_{L}}Z_{2il}^{\ast D_{L}}(%
\overline{s}P_{R}b)(\overline{\tau }P_{L}\mu )\Big]\Big)
\end{eqnarray}
Therefore, one can write the neutralino amplitude as%
\be%
A_{\widetilde{\chi }^{0}}=A_{\widetilde{\chi}^{0}}^{S-P}+A_{\widetilde{\chi }^{0}}^{V-A} , %
\ee%
with
\begin{eqnarray}
A_{\widetilde{\chi }^{0}}^{S-P} &=&-\frac{i}{16\pi ^{2}}\Big(%
(N_{SRR}+N_{SRL})(\overline{s}P_{R}b)\overline{\tau }\mu +(N_{SLL}+N_{SLR})(%
\overline{s}P_{L}b)\overline{\tau }\mu  \nonumber\\
&&+(N_{SRR}-N_{SRL})(\overline{s}P_{R}b)\overline{\tau }\gamma
^{5}\mu +(N_{SLR}-N_{SLL})(\overline{s}P_{L}b)\overline{\tau }\gamma
^{5}\mu \Big),
\end{eqnarray}
and
\begin{eqnarray}
A_{\widetilde{\chi }^{0}}^{V-A} &=&-\frac{i}{16\pi ^{2}}\Big(%
(N_{VLL}+N_{VLR})(\overline{s}\gamma ^{\mu }P_{L}b)(\overline{\tau
}\gamma
_{\mu }\mu )+(N_{VRR}+N_{VRL})(\overline{s}\gamma ^{\mu }P_{R}b)(\overline{%
\tau }\gamma _{\mu }\mu ) \nonumber\\
&&+(N_{VLR}-N_{VLL})(\overline{s}\gamma ^{\mu }P_{L}b)(\overline{\tau }%
\gamma _{\mu }\gamma ^{5}\mu )+(N_{VRR}-N_{VRL})(\overline{s}\gamma
^{\mu }P_{R}b)(\overline{\tau }\gamma _{\mu }\gamma ^{5}\mu )\Big).
\end{eqnarray}
where %
\bea %
N_{SLL} &=&\frac{g^{4}}{2m_{W}^{2}}\Big[\Big((N_{l2}^{\ast
}+\tan \theta
_{W}N_{l1})\Gamma _{2j}^{L_{L}}-N_{l3}^{\ast }\Gamma _{2j}^{L_{R}}\frac{%
m_{\mu }}{m_{W}\cos \beta }\Big)\Big((2\tan \theta _{W}N_{k1}^{\ast
}\Gamma
_{3j}^{L_{R}\ast } \nonumber\\
&&+N_{k3}^{\ast }\Gamma _{3j}^{L_{L}\ast }\frac{m_{\tau }}{m_{W}\cos \beta }%
\Big)\Big((N_{k2}^{\ast }+\frac{1}{2}\tan \theta _{W}N_{k1}^{\ast
})\Gamma _{3i}^{D_{L}}+N_{k3}^{\ast }\Gamma
_{3i}^{D_{R}}\frac{m_{b}}{m_{W}\cos \beta
}\Big) \nonumber\\
&&\times \Big(\frac{2}{3}\tan \theta _{W}N_{l1}^{\ast }\Gamma
_{2i}^{D_{R}\ast }+N_{l3}^{\ast }\Gamma _{2i}^{D_{L}\ast }\frac{m_{s}}{%
m_{W}\cos \beta }\Big)\Big]\frac{m_{W}^{2}m_{\widetilde{\chi }_{l}^{0}}m_{%
\widetilde{\chi }_{k}^{0}}}{8}B(m_{\widetilde{\chi }_{k}^{0}},m_{\widetilde{%
\chi }_{l}^{0}},m_{\widetilde{b}_{i}},m_{\widetilde{l}_{j}}), %
\eea%
\bea%
N_{SRR} &=&\frac{g^{4}}{2m_{W}^{2}}\Big[\Big((N_{k2}+\tan \theta
_{W}N_{k1})\Gamma _{3j}^{L_{L}\ast }-N_{k3}\Gamma _{3j}^{L_{R}\ast }\frac{%
m_{\tau }}{m_{W}\cos \beta }\Big)\Big((2\tan \theta _{W}N_{l1}\Gamma
_{2j}^{L_{R}} \nonumber\\
&&+N_{l3}\Gamma _{2j}^{L_{L}}\frac{m_{\mu }}{m_{W}\cos \beta }\Big)\Big(%
\frac{2}{3}\tan \theta _{W}N_{k1}\Gamma _{3i}^{D_{R}}+N_{l3}\Gamma
_{3i}^{D_{L}}\frac{m_{b}}{m_{W}\cos \beta }\Big) \nonumber\\
&&\times \Big((N_{l2}+\frac{1}{2}\tan \theta _{W}N_{l1})\Gamma
_{2i}^{D_{L}\ast }+N_{l3}\Gamma _{2i}^{D_{R}\ast
}\frac{m_{s}}{m_{W}\cos
\beta }\Big)\Big]\frac{m_{W}^{2}m_{\widetilde{\chi }_{l}^{0}}m_{\widetilde{%
\chi }_{k}^{0}}}{8}B(m_{\widetilde{\chi }_{k}^{0}},m_{\widetilde{\chi }%
_{l}^{0}},m_{\widetilde{b}_{i}},m_{\widetilde{l}_{j}}),%
\eea%
\bea%
N_{SLR} &=&\frac{g^{4}}{2m_{W}^{2}}\Big[\Big((N_{l2}^{\ast
}+\tan \theta
_{W}N_{l1}^{\ast })\Gamma _{2j}^{L_{L}}-N_{l3}^{\ast }\Gamma _{2j}^{L_{R}}%
\frac{m_{\mu }}{m_{W}\cos \beta }\Big)\Big((N_{k2}+\tan \theta
_{W}N_{k1})\Gamma _{3j}^{L_{L}\ast } \nonumber\\
&&-N_{k3}\Gamma _{3j}^{L_{R}\ast }\frac{m_{\tau }}{m_{W}\cos \beta }\Big)%
\Big(\frac{2}{3}\tan \theta _{W}N_{k1}\Gamma
_{3i}^{D_{R}}+N_{l3}\Gamma
_{3i}^{D_{L}}\frac{m_{b}}{m_{W}\cos \beta }\Big) \nonumber\\
&&\times \Big(\frac{2}{3}\tan \theta _{W}N_{l1}^{\ast }\Gamma
_{2i}^{D_{R}\ast }+N_{l3}^{\ast }\Gamma _{2i}^{D_{L}\ast }\frac{m_{s}}{%
m_{W}\cos \beta }\Big)\Big]\frac{m_{W}^{2}}{8}S(m_{\widetilde{\chi }%
_{k}^{0}},m_{\widetilde{\chi }_{l}^{0}},m_{\widetilde{b}_{i}},m_{\widetilde{l%
}_{j}}),%
\eea%
\bea%
N_{SRL} &=&\frac{g^{4}}{2m_{W}^{2}}\Big[\Big((2\tan
\theta _{W}N_{l1}\Gamma
_{2j}^{L_{R}}+N_{l3}\Gamma _{2j}^{L_{L}}\frac{m_{\mu }}{m_{W}\cos \beta }%
\Big)\Big((2\tan \theta _{W}N_{k1}^{\ast }\Gamma _{3j}^{L_{R}\ast
}+N_{k3}^{\ast }\Gamma _{3j}^{L_{L}\ast }\frac{m_{\tau }}{m_{W}\cos \beta }%
\Big) \nonumber\\
&&\Big((N_{k2}^{\ast }+\frac{1}{2}\tan \theta _{W}N_{k1}^{\ast
})\Gamma _{3i}^{D_{L}}+N_{k3}^{\ast }\Gamma
_{3i}^{D_{R}}\frac{m_{b}}{m_{W}\cos \beta
}\Big) \nonumber\\
&&\times \Big((N_{l2}+\frac{1}{2}\tan \theta _{W}N_{l1})\Gamma
_{2i}^{D_{L}\ast }+N_{l3}\Gamma _{2i}^{D_{R}\ast
}\frac{m_{s}}{m_{W}\cos
\beta }\Big)\Big]\frac{m_{W}^{2}}{8}S(m_{\widetilde{\chi }_{k}^{0}},m_{%
\widetilde{\chi
}_{l}^{0}},m_{\widetilde{b}_{i}},m_{\widetilde{l}_{j}}),%
\eea

and \bea%
N_{VLL} &=&\frac{g^{4}}{2m_{W}^{2}}\Big[\Big((N_{l2}^{\ast }+\tan
\theta
_{W}N_{l1})\Gamma _{2j}^{L_{L}}-N_{l3}^{\ast }\Gamma _{2j}^{L_{R}}\frac{%
m_{\mu }}{m_{W}\cos \beta }\Big)\Big((2\tan \theta _{W}N_{k1}^{\ast
}\Gamma
_{3j}^{L_{R}\ast } \nonumber\\
&&+N_{k3}^{\ast }\Gamma _{3j}^{L_{L}\ast }\frac{m_{\tau }}{m_{W}\cos \beta }%
\Big)\Big((N_{k2}^{\ast }+\frac{1}{2}\tan \theta _{W}N_{k1}^{\ast
})\Gamma _{3i}^{D_{L}}+N_{k3}^{\ast }\Gamma
_{3i}^{D_{R}}\frac{m_{b}}{m_{W}\cos \beta
}\Big) \nonumber\\
&&\times \Big(\frac{2}{3}\tan \theta _{W}N_{l1}^{\ast }\Gamma
_{2i}^{D_{R}\ast }+N_{l3}^{\ast }\Gamma _{2i}^{D_{L}\ast }\frac{m_{s}}{%
m_{W}\cos \beta }\Big)\Big]\frac{m_{W}^{2}}{16}S(m_{\widetilde{\chi }%
_{k}^{0}},m_{\widetilde{\chi }_{l}^{0}},m_{\widetilde{b}_{i}},m_{\widetilde{l%
}_{j}}), %
\eea%
\bea%
N_{VRR} &=&\frac{g^{4}}{2m_{W}^{2}}\Big[\Big((N_{k2}+\tan
\theta
_{W}N_{k1})\Gamma _{3j}^{L_{L}\ast }-N_{k3}\Gamma _{3j}^{L_{R}\ast }\frac{%
m_{\tau }}{m_{W}\cos \beta }\Big)\Big((2\tan \theta _{W}N_{l1}\Gamma
_{2j}^{L_{R}} \nonumber\\
&&+N_{l3}\Gamma _{2j}^{L_{L}}\frac{m_{\mu }}{m_{W}\cos \beta }\Big)\Big(%
\frac{2}{3}\tan \theta _{W}N_{k1}\Gamma _{3i}^{D_{R}}+N_{l3}\Gamma
_{3i}^{D_{L}}\frac{m_{b}}{m_{W}\cos \beta }\Big) \nonumber\\
&&\times \Big((N_{l2}+\frac{1}{2}\tan \theta _{W}N_{l1})\Gamma
_{2i}^{D_{L}\ast }+N_{l3}\Gamma _{2i}^{D_{R}\ast
}\frac{m_{s}}{m_{W}\cos
\beta }\Big)\Big]\frac{m_{W}^{2}}{16}S(m_{\widetilde{\chi }_{k}^{0}},m_{%
\widetilde{\chi
}_{l}^{0}},m_{\widetilde{b}_{i}},m_{\widetilde{l}_{j}}),%
\eea%
\bea%
N_{VLR} &=&\frac{g^{4}}{2m_{W}^{2}}\Big[\Big((N_{l2}^{\ast }+\tan
\theta
_{W}N_{l1}^{\ast })\Gamma _{2j}^{L_{L}}-N_{l3}^{\ast }\Gamma _{2j}^{L_{R}}%
\frac{m_{\mu }}{m_{W}\cos \beta }\Big)\Big((N_{k2}+\tan \theta
_{W}N_{k1})\Gamma _{3j}^{L_{L}\ast } \nonumber\\
&&-N_{k3}\Gamma _{3j}^{L_{R}\ast }\frac{m_{\tau }}{m_{W}\cos \beta }\Big)%
\Big(\frac{2}{3}\tan \theta _{W}N_{k1}\Gamma
_{3i}^{D_{R}}+N_{l3}\Gamma
_{3i}^{D_{L}}\frac{m_{b}}{m_{W}\cos \beta }\Big) \nonumber\\
&&\times \Big(\frac{2}{3}\tan \theta _{W}N_{l1}^{\ast }\Gamma
_{2i}^{D_{R}\ast }+N_{l3}^{\ast }\Gamma _{2i}^{D_{L}\ast }\frac{m_{s}}{%
m_{W}\cos \beta }\Big)\Big]\frac{m_{W}^{2}m_{\widetilde{\chi }_{l}^{0}}m_{%
\widetilde{\chi }_{k}^{0}}}{16}B(m_{\widetilde{\chi }_{k}^{0}},m_{\widetilde{%
\chi }_{l}^{0}},m_{\widetilde{b}_{i}},m_{\widetilde{l}_{j}}),%
\eea%
\bea%
N_{VRL} &=&\frac{g^{4}}{2m_{W}^{2}}\Big[\Big((2\tan \theta
_{W}N_{l1}\Gamma
_{2j}^{L_{R}}+N_{l3}\Gamma _{2j}^{L_{L}}\frac{m_{\mu }}{m_{W}\cos \beta }%
\Big)\Big((2\tan \theta _{W}N_{k1}^{\ast }\Gamma _{3j}^{L_{R}\ast
}+N_{k3}^{\ast }\Gamma _{3j}^{L_{L}\ast }\frac{m_{\tau }}{m_{W}\cos \beta }%
\Big) \nonumber\\
&&\Big((N_{k2}^{\ast }+\frac{1}{2}\tan \theta _{W}N_{k1}^{\ast
})\Gamma _{3i}^{D_{L}}+N_{k3}^{\ast }\Gamma
_{3i}^{D_{R}}\frac{m_{b}}{m_{W}\cos \beta
}\Big) \nonumber\\
&&\times \Big((N_{l2}+\frac{1}{2}\tan \theta _{W}N_{l1})\Gamma
_{2i}^{D_{L}\ast }+N_{l3}\Gamma _{2i}^{D_{R}\ast
}\frac{m_{s}}{m_{W}\cos
\beta }\Big)\Big]\frac{m_{W}^{2}m_{\widetilde{\chi }_{l}^{0}}m_{\widetilde{%
\chi }_{k}^{0}}}{16}B(m_{\widetilde{\chi }_{k}^{0}},m_{\widetilde{\chi }%
_{l}^{0}},m_{\widetilde{b}_{i}},m_{\widetilde{l}_{j}}).%
\eea%

Thus, one finds
\begin{eqnarray}
A_{\widetilde{\chi }^{0}}^{S-P}&=&i\frac{f_{B_{s}}N_{F}}{2\left\vert
V_{tb}^{CKM}V_{ts}^{CKM\ast
}\right\vert }\Big(\frac{m_{B_{s}}^{2}}{m_{b}+m_{s}}(R_{0S}-R_{0S}^{\prime })%
\overline{\tau }\mu
+\frac{m_{B_{S}}^{2}}{m_{b}+m_{s}}(R_{0P}-R_{0P}^{\prime
})\overline{\tau }\gamma ^{5}\mu \Big)
\end{eqnarray}
and
\begin{eqnarray}
A_{\widetilde{\chi }^{0}}^{V-A}=i\frac{f_{B_{s}}N_{F}}{2\left\vert
V_{tb}^{CKM}V_{ts}^{CKM\ast }\right\vert }\Big((R_{0V}-R_{0V}^{\prime })(%
\overline{\tau }\gamma _{\mu }\mu )+(R_{0A}-R_{0A}^{\prime
})(\overline{\tau }\gamma _{\mu }\gamma ^{5}\mu )\Big)
\end{eqnarray}
where
\begin{eqnarray}
R_{0V,A} &=&\frac{i}{16\pi ^{2}}\frac{\left\vert
V_{tb}^{CKM}V_{ts}^{CKM\ast
}\right\vert }{N_{F}}(N_{VLR}\pm N_{VLL}) \\
R_{0V,A}^{\prime } &=&\frac{i}{16\pi ^{2}}\frac{\left\vert
V_{tb}^{CKM}V_{ts}^{CKM\ast }\right\vert }{N_{F}}(N_{VRR}\pm N_{VRL}) \\
R_{0S,P} &=&\frac{i}{16\pi ^{2}}\frac{\left\vert
V_{tb}^{CKM}V_{ts}^{CKM\ast
}\right\vert }{N_{F}}(N_{SRR}\pm N_{SRL}) \\
R_{0S,P}^{\prime } &=&\frac{i}{16\pi ^{2}}\frac{\left\vert
V_{tb}^{CKM}V_{ts}^{CKM\ast }\right\vert }{N_{F}}(N_{SLR}\pm
N_{SLL}).
\end{eqnarray}
In this case, these co-efficients can be used in Eq.~\ref{BrBlilj} to calculate the branching ratio for
$B_{s}\rightarrow \mu^{+}\tau ^{-}$.

\subsection{Numerical Analysis}

We can estimate the typical prediction of SUSY models to the
branching ration of the $B_s \to \tau \bar{\mu} + \mu \bar{\tau}$
process. As usual one can easily show that the neutralino
contribution is typically one or two order of magnitude less than
the light chargino effect. Therefore, here we will neglect the
neutralino diagrams and focus on the dominate chargino contribution.
As explicitly shown above the chargino amplitude depends on stops
and tau-sneutrinos masses and mixing. Therefore a light stop and
light tau-sneutrinos may significantly enhance the chargino
contribution to $BR(B_s \to \tau \bar{\mu} + \mu \bar{\tau})$.

In MSSM, the stop mass matrix is given by%
\be%
M^2_{\tilde{t}}= \left(\begin{array}{cc} m_{Q}^2 + m_t^2 +
\frac{1}{6}M_Z^2\cos2\beta(3-4\sin^2\theta_W)
&m_t(A_t-\mu \cot\beta )\\
m_t(A_t-\mu\cot\beta)& m_{U}^2 + m_t^2 + \frac{2}{3}M_Z^2\cos
2\beta\sin^2\theta_W\end{array}\right), %
\ee%
where $m_{Q}^2$ and $m_{U}^2$ are the low energy values of the
soft SUSY breaking scalar masses. Due to large off-diagonal
elements, the diagonalization of the stop mass-squared matrix
leads to the physical eigenstates with one light and one heavy
stops. In addition, with no right-handed neutrinos in MSSM the
sneutrino mass matrix is given by %
\be%
M^2_{\tilde{\nu}} = M_L^2 + \frac{1}{2} M_Z^2 \cos^2 \beta ,%
\ee%
where $M_L$ is the soft SUSY scalar masses of the sleptons. It is
also a part of the left-left sector of the charged slepton
mass-squared matrix. However, it is worth noting that the
experimental limits of lepton flavor violations, like $\mu \to e
\gamma$ impose stingiest constraints on the off-diagonal elements
of $M_L^2$ between the first and second generations, while the
mixing between first or the second and third generations are much
less constrained \cite{Gabbiani:1996hi}.

\begin{figure}[t]
\begin{center}
\epsfig{file=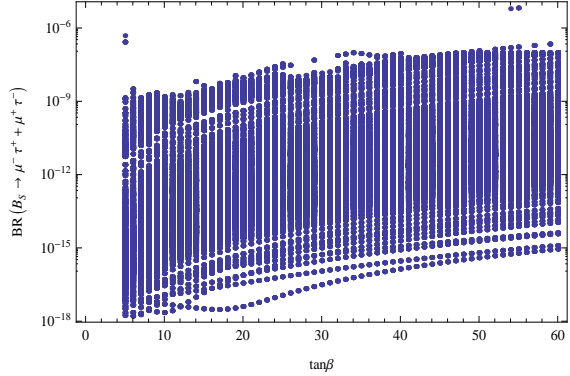,height=5.0cm,width=7.5cm,angle=0}
\vspace{-0.5cm}
\end{center}
\caption{ The chargino contribution for $BR(B_s \to \tau \bar{\mu} +
\mu \bar{\tau})$ as function of $\tan \beta$ for stop masses of
order $200$ and $800$ GeV and $M_2 \in[100,1000]$ GeV and $\mu \in
[250,1000]$ GeV.} \label{hgg}
\end{figure}

In Fig. 4 we present the $BR(B_s \to \tau \bar{\mu} + \mu
\bar{\tau})$ as function of $\tan \beta$. We assume that the
lightest and heaviest stop masses are of order 200 GeV and $800$ GeV
respectively. We consider $100~ {\rm GeV} \leq M_2 \leq 1~ {\rm
TeV}$ and take $\mu$ between 250 GeV and 1 TeV. The lightest
sneutrino mass is assumed to be of order 200 GeV and the sneutrino
mixing parameters are of order ${\cal O}(0.1)$. As can be seen from
this figure, the branching ratio $BR(B_s \to \tau \bar{\mu} + \mu
\bar{\tau})$ can be enhanced up to $10^{-7}$ for large $\tan \beta$
and light charginos and sneutrinos. Also if we have semi-degenerate
masses for chargino, stops, and sneutrino, then a kind of a
resonance in the branching ratio may be obtained (as can be seen
from the loop
functions: $B(m_{\widetilde{\chi }_{k}^{+}},m_{\widetilde{\chi }_{l}^{+}},m_{\widetilde{t%
}_{i}},m_{\widetilde{\nu }_{j}})$ and $S(m_{\widetilde{\chi }_{k}^{+}},m_{\widetilde{\chi }_{l}^{+}},m_{\widetilde{t%
}_{i}},m_{\widetilde{\nu }_{j}})$). In this case the branching ratio
can reach $10^{-6}$, however, we think this is a fine tuning
therefore we assume that the mass difference between these particles
$> 50$ GeV at least. From this analysis one can conclude that the
typical SUSY predictions for this branching ratio of $B_s \to \tau
\bar{\mu} + \mu \bar{\tau}$ is of order $10^{-8}-10^{-9}$.

Finally, it is worth noting that in the second paper of
Ref.\cite{btaumu_old}, it was emphasized that in the supersymmetric
seesaw extensions of the SM, the two loop double penguin diagrams may
be relevant at a very large $\tan \beta$ (of order 60) even with
flavor universal slepton masses. In this case, it was shown that
the double penguins mediated by Higgs bosons may induce a
branching ratio of order $B_s \to \mu^+ \tau^- \sim 10^{-7} \times
(\tan \beta/60)^8 \times (100~{\rm GeV}/m_A)^4$, which may give
dominant results at very large $\tan \beta$ and low $m_A$.

\section{Summary}

In this paper we analyzed the rare decay $B \to \tau \bar{\mu} + \mu \bar{%
\tau}$ in two possible extensions of the SM. In particular, we considered
the
decay in a model with an extended higgs sector, where the decay is mediated by a light
scalar which is identified with the recently observed state at LHC. In the
second example, the one loop contribution to this process in MSSM was
computed. We found that in the former scenario the branching ratio $B_s \to
\tau \bar{\mu} + \mu \bar{\tau}$ can be as high as $10^{-6}$ ( $B_d \to \tau \bar{%
\mu} + \mu \bar{\tau} $ can be as high as $ \sim 10^{-7}$), while in
the later scenario the $BR(B_s \to \tau \bar{\mu} + \mu \bar{\tau})
$ could be  $ \sim 10^{-8}$. Therefore, one concludes that probing
this lepton flavor violating process at the LHCb would be a
significant hint for a non-traditional new physics beyond the SM
with large FCNC couplings.

\bigskip

\hspace{-0.4cm}\textbf{Acknowledgments:}

D.B. is partially supported by the Algerian Ministry of Higher Education and
Scientific Research under the PNR 'Particle Physics/Cosmology: the
interface', and the CNEPRU Project No. D01720090023. S.K. thanks The
Leverhulme Trust (London, UK) for financial support in the form of a
Visiting Professorship to the University of Southampton. The work of A.D.
and M.D. was financially supported by the US-Egypt Joint Board on Scientific
and Technological Co-operation award (Project ID: 1855) administered by the
US Department of Agriculture.

\end{document}